\documentclass[10pt]{article}
\usepackage{graphicx} \usepackage{amssymb}
\newcommand{\Tr}{\hbox{Tr}}
\newcommand{\ket}[1]{\vert #1 \rangle}

\newcommand{\ketbra}[2]{\vert #1 \rangle \! \langle #2 \vert}

\begin{document}
\title{\bf About distillability of depolarized states}
\author{Andrea R. Rossi${}^{1,2}$ and Matteo G. A. Paris${}^{1,3}$\\
${}^1$Dipartimento di Fisica, Universit\`a di Milano, Italia\\
${}^2$INFM, Unit\`a di Milano, Italia \\
${}^3$Dipartimento di Fisica "A. Volta", Universit\`a di Pavia,
Italia}
\date{}
\maketitle
%%%%%%%%%%%%%%%%%%%%%%%%%%%%%%%%%%%%%%%%%%%%%%%%%%%%%%%%%%%%%%%%%%%
\begin{abstract}
Reduction criteria for distillability is applied to general
depolarized states and an explicit condition is found in terms
of a characteristic polynomial of the density matrix. 3
$\times$ 3 bipartite systems are analyzed in some details.
\end{abstract}
%%%%%%%%%%%%%%%%%%%%%%%%%%%%%%%%%%%%%%%%%%%%%%%%%%%%%%%%%%%%%%%%
\section{Introduction}
Quantum information is mostly based on use of entanglement, and
optimal performances of quantum protocols are obtained when the
parties share maximally entangled states. However, entanglement is
corrupted by the interaction with the environment, and it becomes
crucial to establish whether or not quantum protocols may be
exploited for a given level of the environmental noise. Quantum
purification consists \cite{dhb} in distilling a certain number of
maximally entangled states from a larger number of corrupted
states, {\em i.e.} to sacrifice some copies of the received state
to increase the entanglement of others. Entanglement of partially
corrupted states is only a necessary conditions for distillation,
{\em bound entanglement} being the entanglement that cannot be
distilled \cite{hf}. Some sufficient criterion for distillability
has been proposed, the most relevant being the Reduction Criterion
(RC) \cite{horo1}, and the norm-related criterions \cite{norm}.
\par
In this paper we apply the RC to a generalized class of
depolarized states in $d$ dimensions. First we
briefly review the distillability criterion and introduce a
characteristic polynomial for our class of states. Then we apply
our results to $3 \times 3$ bypartite systems and numerically study
the distillability.
%%%%%%%%%%%%%%%%%%%%%%%%%%%%%%%%%%%%%%%%%%%%%%%%%%%%%%%%%%%%%%%%%%%%%%%%%%%%%%%
\section{The Reduction Criterion} \label{rc}
We consider $d\times d$ bipartite systems. RC for
distillability is based on map theory \cite{horo1} and states
that the negativity of the matrix
\begin{eqnarray} \label{red:crit}
\rho_{A}\otimes I-\rho\leq0,
\end{eqnarray}
where $\rho$ is the systems's density operator and $\rho_{A}=
\Tr_{\!B}{\rho}$ is the reduced density operator, implies the
distillability of the original density matrix $\rho$. Using RC we
will study the distillability of depolarized states of the form
\begin{eqnarray} \label{our:state}
\rho=p\,\ketbra{\psi}{\psi}+\frac{1-p}{d^{2}}\,\mathbb{I}\:,
\end{eqnarray}
where $\ket{\psi}=\sum_{i}{a_{i}\ket{ii}}$ is a generic pure state.
We already considered the Schmidt's decomposition of $|\psi\rangle$
and thus, without loss of generality, the $a_{i}$'s are supposed to
be real numbers, with the additional condition $\sum_{i}a_{i}^{2}=1$
\cite{schmidt}. $\mathbb{I}=\sum_{i,j}\ketbra{ij}{ji}$ denotes the
$d \times d$ identity matrix. Depolarized maximally entangled state
are obtained for $\ket{\psi}=\sum_{i}{\frac{1}{\sqrt{d}}\ket{ii}}$
\par
To make the RC (\ref{red:crit}) of some computational use, we have
to explicitly write down the $d^2 \times d^2$ dimensional real
matrix representing (\ref{our:state}). In order to do this we
first find out the Fock representation of (\ref{red:crit}) if
$\rho$ is given by (\ref{our:state}). We have:
\begin{eqnarray} \label{fock:red:crit}
\rho_{A} \otimes \mathbb{I}-\rho=\sum_{i,j} \left\{ p\left(
\,a_{i}^{2}\ketbra{ij}{ji}-a_{i}a_{j}\ketbra{ii}{jj}\right)+\frac{(d-1)(1-p)}{d^{2}}\ketbra{ij}{ji}
\right\}
\end{eqnarray}
or, in matrix form:
\begin{eqnarray} \label{matr:red:crit}
\rho_{A} \otimes \mathbb{I}-\rho\,=\,p\,\left( \begin{array}{cccc}
\mathbf{A}_{11} & \mathbf{A}_{12} &
\cdots & \mathbf{A}_{1d} \\
\mathbf{A}_{21} & \mathbf{A}_{22} & \cdots & \mathbf{A}_{2d} \\
\vdots & & \ddots & \vdots \\
\mathbf{A}_{d1} & \cdots & \cdots & \mathbf{A}_{dd}
\end{array} \right)+ f_{d}(p) \, \mathbb{I}_{d^{2} \times d^{2}},
\end{eqnarray}
where the $\mathbf{A}$'s are $d\times d$ matrices given by
\begin{eqnarray} \label{Aii}
\mathbf{A}_{ii} = a_i^2 \left(\mathbb{I}-\mathbb{D}[i,i]\right)
\qquad \mathbf{A}_{ij} = - a_i a_j \mathbb{D}[i,j]
\qquad \mathbf{A}_{ji}=\mathbf{A}_{ij}^{\mathrm{T}}
\label{Aij}\:,
\end{eqnarray}
and $f_{d}(p)\,=\,\frac{(d-1)(1-p)}{d^{2}}$. In Eqs. (\ref{Aij})
$\mathbb{D}[i,j]$ denotes a matrix which is zero everywhere except
for the $(i,j)$ entry, which is one. We are now in the position to
write the characteristic polynomial $\mathbb{P}_d$ of the matrix
(\ref{matr:red:crit}) as a product of two terms
\begin{eqnarray} \label{full:poly}
\mathbb{P}_d =\mathrm{P}_{d}(x)\,\prod_{i=1}^{d}\left(
f_{d}(p)-\lambda+p\,a_{i}^{2}\right)^{d-1}\:,
\end{eqnarray}
where, using the substitution $x=f_{d}(p)-\lambda$, we have
\begin{eqnarray} \label{intr:poly}
\mathrm{P}_{d}(x)\,=\,x^d - \sum_{i = 0}^{d - 2}
      \left( d - i - 1 \right) \,p^{d - i}\,A_{d-i}^{d}\,x^i\:,
\end{eqnarray}
with
\begin{eqnarray}
A_{k}^{d}=\sum_{j = 0}^{d - k}
A_{k-1}^{d-j-1}\,\left( A_{1}^{d-j} - A_{1}^{d-j-1} \right)
\qquad A_{1}^{d} = \sum_{l=1}^{d}a_{l}^{2}\:. \label{subst}
\end{eqnarray}
Notice that the eigenvalues given by the second factor in
(\ref{full:poly}) are always positive. Therefore, the sign of
$\mathbb{P}_d$ is determined by (\ref{intr:poly}). Notice also
that in (\ref{intr:poly}) the term $x^{d-1}$ is always missing.
%%%%%%%%%%%%%%%%%%%%%%%%%%%%%%%%%%%%%%%%%%%%%%%%%%%%%%%%%%%%%%%%%%%%%%%%%%%%
\subsection{Miscellaneous limits}
As a first application we consider the case $a_{i}=d^{-1/2}$,
\emph{e.g.} when $\ket{\psi}$ is a maximally entangled state. The
polynomial (\ref{intr:poly}) reduces to
\begin{eqnarray} \label{max:ent:poly}
\mathrm{P}_{d}(x)|_{a_{i}=d^{-1/2}}=-
\frac{1}{d^{d}}(p+d\,x)^{d-1}((d-1)p-d\,x).
\end{eqnarray}
The only relevant eigenvalue is the one given by the last factor
of the $\mathrm{P}$ polynomial, namely
\begin{eqnarray} \label{eigen:max:ent}
\lambda=-\frac{1-d}{d^{2}}+\frac{1-d^{2}}{d^{2}}\,p,
\end{eqnarray}
that yields the already known condition \cite{horo1}
$p\geq\frac{1}{d+1}$ for distillability of
depolarized maximally entangled states.
\par
As a second check consider the case of a $(d-j)$-dimensional
maximally entangled state in a $d$-dimensional space, {\em i.e.}
$a_i=1/\sqrt{(d-j)}$, $i=1,...,d-j$ while the remaining $j<d$
coefficients $a_{i}$ are set to zero. In this case it's not
difficult to see that the $d$-dimensional $\mathrm{P}$ polynomial
(\ref{intr:poly}) reduces to the $(d-j)$-dimensional one
\begin{eqnarray} \label{max:ent:minus}
\mathrm{P}_{d-j}(x)
=\,-(d-j)^{-(d-j)}(p+(d-j)x)^{d-j-1}((d-j-1)p-(d-j)x)\:.
\end{eqnarray}
The non-trivial root of (\ref{max:ent:minus}) reads
\begin{eqnarray} \label{eigen:max:ent:minus}
x=f_{d}(p)-\lambda=\frac{d-j-1}{d-j}\:,
\end{eqnarray}
leading to the condition
\begin{eqnarray} \label{cond:max:ent:minus}
p\geq\frac{(d-1)(d-j)}{(d^{2}-1)(d-j)-d\,j}\:.
\end{eqnarray}
For example, for a $3\times 3$ system and a state with one
zero-coefficient we find that the distillability is assured for
$p>4/13$. Notice that RC is only a sufficient condition for
distillability and therefore the bound in
(\ref{cond:max:ent:minus}) need not to be a lower bound. Indeed,
for the $3\times 3$ case considered above the system is also
distillable for $p>2/11$ \cite{vidal}.
%%%%%%%%%%%%%%%%%%%%%%%%%%%%%%%%%%%%%%%%%%%%%%%%%%%%%%%%%%%%%%%%%%%%%%%%%%%%
\section{The $3\times 3$ dimensional bipartite system} \label{3x3sys}
\begin{figure}[ht]
\begin{center}
\includegraphics[width=0.75\textwidth]{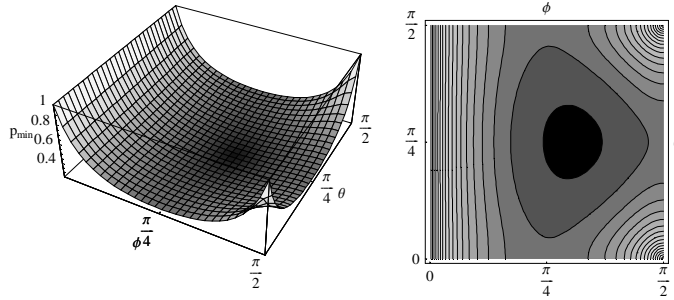}
\caption{3D and contour plots of the minimum $p$ satisfying RC
as a function of state's coefficients written in polar coordinates.
}\label{f:qbed}
\end{center}
\end{figure}

As a further example we apply this representation to the case of a
$3 \times 3$ bipartite system. In this section, since $d$ is set
to 3, we rename $A_{i}^{3} \equiv B_{i}$. Equation
(\ref{matr:red:crit}) becomes:
\begin{eqnarray} \label{3x3:mat}
\rho_{A} \otimes \mathbb{I}-\rho\,&=&\,p\,\left(
\begin{array}{ccccccccc}
0 & 0 & 0 & 0 & -a_{1}a_{2} & 0 & 0 & 0 & -a_{1}a_{3} \\
0 & a_{1}^{2} & 0 & 0 & 0 & 0 & 0 & 0 & 0 \\
0 & 0 & a_{1}^{2} & 0 & 0 & 0 & 0 & 0 & 0 \\
0 & 0 & 0 & a_{2}^{2} & 0 & 0 & 0 & 0 & 0 \\
-a_{1}a_{2} & 0 & 0 & 0 & 0 & 0 & 0 & 0 & -a_{2}a_{3} \\
0 & 0 & 0 & 0 & 0 & a_{2}^{2} & 0 & 0 & 0 \\
0 & 0 & 0 & 0 & 0 & 0 & a_{3}^{2} & 0 & 0 \\
0 & 0 & 0 & 0 & 0 & 0 & 0 & a_{3}^{2} & 0 \\
-a_{1}a_{3} & 0 & 0 & 0 & -a_{2}a_{3} & 0 & 0 & 0 & 0
\end{array} \right) \nonumber \\
\nonumber \\
&\hbox{}& +\frac{2}{9}\,(1-p)\,\mathbb{I}_{9 \times 9}.
\end{eqnarray}
The  polynomial (\ref{intr:poly}) now reads:
\begin{eqnarray} \label{3x3:poly}
\mathrm{P}_{3}(x)\,=\,x^{3}-(d-2)\,p^2\,B_{2}\,x-(d-1)\,p^3\,B_{3}
\end{eqnarray}
with
\begin{eqnarray} \begin{array}{lcr} \label{coeff:3x3:poly}
B_{2}=a_{1}^{2}a_{2}^{2} + a_{1}^{2}a_{3}^{2} + a_{2}^{2}a_{3}^{2}
&\phantom{0.5cm} &B_{3}=a_{1}^{2}a_{2}^{2}a_{3}^{2}
\end{array}
\end{eqnarray}
The solutions of (\ref{3x3:poly}) are:
\begin{eqnarray}
x_{1}&=&2\,p\,\sqrt{\frac{B_{2}^{3}}{3}}\,\cos\left[\frac{1}{3}\,
\arctan\left(\sqrt{\frac{B_{2}^{3}}{27\,B_{3}^{2}}-1}
\right)\right] \label{sol:3x3:1} \\
x_{\pm}&=&-p\,\sqrt{\frac{B_{2}^{3}}{3}}\,\left\{\cos\left[
\frac{1}{3}\,\arctan\left(\sqrt{\frac{B_{2}^{3}}{27\,B_{3}^{2}}-1}
\right)\right]\right. \nonumber \\
&\hbox{}&\,\pm\left.\sqrt{3}\,\sin\left[
\frac{1}{3}\,\arctan\left(\sqrt{\frac{B_{2}^{3}}{27\,B_{3}^{2}}-1}
\right)\right]\right\} \label{sol:3x3:23}
\end{eqnarray}
and the nontrivial eigenvalues of (\ref{3x3:mat}) are:
\begin{eqnarray}
\lambda_{i}&=&\frac{2}{9}\,(1-p)-x_{i}\,\,, \,i=1,+,-\:.
\label{eigenvalues}
\end{eqnarray}
We are now in the condition to find out for which values of $p$ at
least one (if any) of the nontrivial eigenvalues is negative. To
this end, we switch to spherical coordinates, setting
$a_{1}\rightarrow\sin\theta\,\sin\phi\ $, $
a_{2}\rightarrow\cos\theta\,\sin\phi\ $, $
a_{3}\rightarrow\cos\phi $, and plot the minimum value of $p$ for
which (\ref{red:crit}) is satisfied, as a function of $\theta$ and
$\phi$. Doing this for every eigenvalue (\ref{eigenvalues}), we
than select the minimum value of $p$ between the three, at any
given point on the plane $(\theta,\phi)$. It turns out that the
only contributing eigenvalue is $\lambda_{1}$. Our results are
plotted in Fig. \ref{f:qbed}.
%%%%%%%%%%%%%%%%%%%%%%%%%%%%%%%%%%%%%%%%%%%%%%%%%%%%%%%%%%%%%%%%%%%%%%%%%%%%
\section{Conclusions} \label{concl}
In this paper we have explicitly written the characteristic
equation coming from the application of the Reduction Criterion
for distillability to a class of one parameter generalized
depolarized $d \times d$ bipartite states. The method has been
applied to $3 \times 3$ bipartite systems and the
minimum value of the parameter $p$ that guarantees distillability has
been found.
\section*{Acknowledgements}
The work of MGAP is partially supported by the EC program ATESIT (Contract
No. IST-2000-29681).
%%%%%%%%%%%%%%%%%%%%%%%%%%%%%%%%%%%%%%%%%%%%%%%%%%%%%%%%%%%%%%%%%%%%%%%%%%%%

%%%%%%%%%%%%%%%%%%%%%%%%%%%%%%%%%%%%%%%%%%%%%%%%%%%%%%%%%%%%%%%%%%%%%%

%%%%%%  end  %%%%%%%%%%%%%%%%%%%%%%%%%%%%%%%%%%%%%%%%%%%%%%%%%%%%%%%%%%%%%%%%%%
\end{document}